\documentclass[aps, prl, twocolumn, nofootinbib]{revtex4}

\usepackage{amsfonts, amssymb, amsmath}
\usepackage{bm, bbm}

\newcommand{\eqn}[1]{$\displaystyle{#1}$}

\newcommand{\eql}[1]{\label{eq:#1}}
\newcommand{\eq}[1]{(\ref{eq:#1})}
\newcommand{\Eq}[1]{eq.$\,$(\ref{eq:#1})}
\newcommand{\nn}{\nonumber}

\newcommand{\gsim}{\gtrsim}

\newcommand{\del}{\partial}
\newcommand{\dee}{\mathrm{d}}

\newcommand{\fr}[2]{\frac{#1}{#2}}

\newcommand{\vev}[1]{\langle#1\rangle}

\newcommand{\De}{\Delta}
\newcommand{\vep}{\varepsilon}
\newcommand{\la}{\lambda}
\newcommand{\La}{\Lambda}
\newcommand{\tha}{\theta}

\newcommand{\cL}{\mathcal{L}}
\newcommand{\cO}{\mathcal{O}}
\newcommand{\cR}{\mathcal{R}}

\newcommand{\Z}{\mathbb{Z}}
\newcommand{\R}{\mathbb{R}}

\newcommand{\Mpl}{M_\text{Pl}}
\newcommand{\cmb}{\text{CMB}}

\begin{document}

\title{Gyroscopic Inflation}

\author{Alejandro Jenkins}\email{jenkins@hep.fsu.edu}
\author{Takemichi Okui}\email{okui@hep.fsu.edu}
\affiliation{Department of Physics, Florida State University, 
Tallahassee, Florida 32306-4350, USA}

\begin{abstract}
We propose a new framework for multi-field inflation in which a nearly constant 
potential energy is maintained during inflation before decreasing rapidly, in a 
manner analogous to a classical top spinning upright for a long time before 
falling down. We provide the simplest realization of such dynamics as a 
well-controlled, weakly-coupled effective field theory with a global shift 
symmetry. Nonperturbative quantum gravitational effects, which break global 
symmetries, are suppressed in this model. Primordial gravitational waves may 
be within experimental reach.
\end{abstract}

\maketitle

%%%%%%%%%%%%%%%%%%%%%%%%%%%%%%%%%%%%%%%%%%%%%%%%%%%%%%%%%%%%%%%%%%%%%%%%%%%%%%%%
%%%%%%%%%%%%%%%%%%%%%%%%%%%%%%%%%%%%%%%%%%%%%%%%%%%%%%%%%%%%%%%%%%%%%%%%%%%%%%%%

%%%%%%%%%%%%%%%%%%%%%%%%%%%%%%%%
\section{Introduction and Model}
%%%%%%%%%%%%%%%%%%%%%%%%%%%%%%%%

The spectacular data from the cosmic microwave background (CMB) and large-scale 
structure measurements have firmly established slow-roll 
inflation~\cite{Linde:1981mu, Albrecht:1982wi} as the source of primordial 
density fluctuations,%
\footnote{See \cite{Baumann:2009ds} for a recent excellent review.}
as well as a solution to the old cosmological conundrums such as the horizon 
problem~\cite{Guth:1980zm}. With the ongoing Planck satellite mission and the 
next generation of CMB experiments, we are entering an exciting era in 
which we will be probing some of the fundamental parameters of inflationary 
dynamics. 

A challenging problem in slow-roll inflation, from the viewpoint of effective 
field theory, is how to protect its extremely flat potential against quantum 
corrections. This problem appears robust, as extreme flatness is required to 
maintain inflation for a sufficiently long time, while the loop corrections can 
be ultimately attributed to the need for inflation to end. If inflation were to 
last forever, we could ensure an exactly flat potential for a slow-roll scalar 
field, \eqn{\psi}, even in the presence of radiative corrections, by imposing 
symmetry under the global shift \eqn{\psi(x) \to \psi(x) + c}. But inflation 
must end, which seems to require violating the shift symmetry to let the 
potential drop to zero at some point, thereby unleashing the dangerous radiative 
corrections. 

In this letter, we propose a new framework that provides a simple, concrete 
counterexample to this argument. We dub the framework ``gyroscopic inflation,'' 
as its dynamics can be essentially captured by analogy with a symmetric top 
spinning on a nearly frictionless, horizontal table. The top's potential 
energy evolves precisely in the manner required for a successful slow-roll 
inflation; it stays approximately constant for a long time while the top remains 
upright and then swiftly drops to zero as the top falls down. Let us highlight 
two essential features of this dynamics in terms of the Euler angles $\tha$ and 
$\psi$ (the top's tilt and its spinning), neglecting the slow precession 
variable $\phi$: (i) $\tha$ retains a nearly constant potential energy not by 
slowly rolling over a very flat potential but by sitting at the origin 
\eqn{\tha = 0}, which remains a stable point for a long time; (ii) The stability 
of the origin is controlled not by how much the top has turned, $\psi$, but by 
how fast it is turning, $\dot{\psi}$, which is invariant under the shift 
\eqn{\psi \to \psi + c}.

To implement these features in relativistic field theory, we promote $\psi$ 
to a slow-roll field, and $\tha$ to a ``waterfall'' field, borrowing the 
terminology of hybrid inflation~\cite{Linde:1991km, Linde:1993cn}, 
which shares the feature (i) but differs crucially in regard to (ii); 
namely, in hybrid inflation, it is $\psi$ itself that controls the 
stability of $\tha$, so the shift symmetry is necessarily violated.%
\footnote{Nevertheless, various mechanisms exist to tame the loop corrections 
in hybrid inflation, e.g., by applying the little higgs 
mechanism~\cite{Kaplan:2003aj, ArkaniHamed:2003mz} or by making the inflaton 
and waterfall fields composite~\cite{Sundrum:2009ii}.}
Now, since $\tha$ should be unstable by itself, we write a tachyonic 
mass-squared 
term, \eqn{+\mu^2 \tha^2}, in the lagrangian. In the presence of a sufficiently 
large~\eqn{|\dot{\psi}|}, \eqn{\tha = 0} should become a stable point. The 
leading $\psi$-$\tha$ cross coupling with such a property is \eqn{- (\del_\mu 
\psi) (\del^\mu \psi) \tha^2 = -\dot{\psi}^2 \tha^2 + (\del_i \psi) (\del^i 
\psi) \tha^2}, which quickly redshifts to \eqn{-\dot{\psi}^2 \tha^2} in an 
expanding spacetime and provides $\tha$ with a positive effective mass-squared. 
Therefore, the simplest model of gyroscopic inflation is given by%
\begin{align}
\cL =
&\, 
\fr12 (\del \psi)^2 - \fr{m^2}{2} \psi^2  \nn\\
&+ \fr12 (\del \tha)^2  
 - \fr12 \biggl[ -\mu^2 + \fr{(\del \psi)^2}{\La^2} \biggr] \tha^2 
 - \fr{\la}{4} \tha^4  \nn\\
&- V_0 + \cL_\text{ct}
\,,\eql{lagrangian}
\end{align}
where $m^2$, $\mu^2$, $\La^2$ and $\la$ are all positive real parameters, and 
$V_0$ adjusts the energy of the vacua (located at \eqn{(\psi, \tha) = \bigl( 0, 
\pm \mu / \sqrt{\la} \bigr)} at tree level) to zero. 
Internal symmetries imposed  on~\eq{lagrangian} are: (a) two $\Z_2$ symmetries 
under which $\tha \to -\tha$ and $\psi \to -\psi$, and (b) a global \eqn{\psi 
\to \psi + c} shift symmetry broken only by the \eqn{m^2 \psi^2} term.%
\footnote{This assumption can be made only in the limit of decoupling gravity, 
\eqn{\Mpl \to \infty}. With gravity turned on, the term \eqn{\sqrt{-g} \, m^2 
\psi^2} yields an infinite number of shift-symmetry breaking terms when expanded 
in terms of the graviton field. The effects of gravity will be discussed below.}
Being nonrenormalizable, 
this theory has a physical momentum cutoff proportional to $\La$, with the 
proportionality constant depending on the ultraviolet completion of the theory.   
$\cL_\text{ct}$ denotes all counterterms that are not only consistent with these 
symmetries but also \emph{required} by renormalization. In particular, 
$\cL_\text{ct}$ is exactly invariant under \eqn{\psi \to \psi + c}, since the 
breaking by \eqn{m^2 \psi^2} is soft and does not introduce any ultraviolet 
divergence that breaks the shift symmetry.%
\footnote{Again, this is true only in the limit of decoupling gravity.}

The \eqn{m^2 \psi^2} term cannot be omitted. An 
inflationary phase lasts as long as \eqn{|\dot{\psi}|} remains large enough to 
keep $\tha$ stable at \eqn{\tha = 0}. As the universe expands, 
\eqn{|\dot{\psi}|} decreases via Hubble damping and inflation thus ends 
eventually, but if $m^2$ were zero, the damping rate of $\dot{\psi}$ would be of 
order the expansion rate $H$ itself by dimensional analysis. This would be too 
fast to give a sufficient number of e-folds. On the other hand, with a nonzero 
but still small \eqn{m^2 \ll H^2}, $\psi$ behaves as a slow-roll field with a 
damping rate \eqn{\sim m^2 / H \ll H}, permitting us to obtain enough e-folds.  

%%%%%%%%%%%%%%%%%%%%%%%%%%%%%%%%%%%%%%%%%%
\section{Dynamics of Gyroscopic Inflation}
%%%%%%%%%%%%%%%%%%%%%%%%%%%%%%%%%%%%%%%%%%

Let us investigate the dynamics of gyroscopic inflation in its simplest 
realization~\eq{lagrangian}. We choose an initial condition in which $\tha$ is 
displaced from a vacuum \eqn{\vev{\tha}} toward the origin, that is, \eqn{\tha^2 
\ll \vev{\tha}^2}. We also assume a sufficiently large initial 
\eqn{|\dot{\psi}|} such that the origin \eqn{\tha = 0} is initially stable. The 
universe will then expand with a rate \eqn{H \gsim \sqrt{V_0 / (3\Mpl^2)}}, 
quickly redshifting away spatial curvature as well as any spatial gradients of 
$\tha$ and $\psi$.  Therefore, in coordinates in which \eqn{\dee s^2 = \dee t^2 
- [a(t)]^2 \dee x^i \dee x^i}, the equation of motions for $\tha$ and $\psi$ 
are given at tree level by%
\begin{align}
\ddot{\tha} + 3H \dot{\tha} 
+ \biggl( \fr{\dot{\psi}^2}{\La^2} - \mu^2 \biggr) \tha 
&= 0  \,,\eql{theta-eom}\\
\ddot{\psi} + 3H \dot{\psi} + m^2 \psi 
&= 0  \,,\eql{psi-eom}
\end{align}
where \eqn{H \equiv \dot{a} / a}, and \eqn{\la \tha^3} is dropped in 
\Eq{theta-eom} in accord with the condition \eqn{\tha^2 \ll \vev{\tha}^2}. 
Also, in deriving \Eq{psi-eom}, the $\psi$-$\tha$ cross-coupling term in the 
lagrangian~\eq{lagrangian} was ignored,%
\footnote{Therefore, unlike in $k$-inflation~\cite{ArmendarizPicon:1999rj}, 
the equation of motion for $\psi$ effectively has a standard kinetic term.}
because $\cL_\text{ct}$ is under control as an expansion in powers of $\tha^2 / 
\La^2$ only if%
\begin{align}
\vev{\tha}^2 \ll \La^2
\,,\eql{vev-less-than-cutoff}
\end{align}
which, together with \eqn{\tha^2 \ll \vev{\tha}^2}, implies \eqn{\tha^2 / \La^2 
\ll 1}. Finally, the inflation rate $H$ in~\eq{theta-eom} and~\eq{psi-eom} is 
given by%
\begin{align}
H^2 = \fr{V_0}{3\Mpl^2}
\,,\eql{expansion-rate}
\end{align}
provided that the energy density of the $\tha$-$\psi$ system is dominated by 
$V_0$, which we will justify shortly. Now, with $H$ being constant, 
\Eq{theta-eom} implies that $\tha$ remains stable at \eqn{\tha = 0} as long as 
\eqn{\dot{\psi}^2 / \La^2 > \mu_\text{c}^2}, where%
\begin{align}
\mu_\text{c}^2
\equiv 
\mu^2 + \biggl( \fr{3H}{2} \biggr)^{\!\!2}
\simeq
\mu^2 
\,.
\end{align}
Here, for a swift ending of inflation, we have assumed \eqn{\mu^2 \gg (3H / 
2)^2} so that $\tha$ is not a slow-roll field and will quickly roll down to one 
of the vacua once it becomes tachyonic. Therefore, at the end of inflation 
(\eqn{t \equiv 0}), we choose \eqn{\dot{\psi}(0) = \pm \mu \La}. Then, 
since $\psi$ \emph{is} a slow-roll field (i.e., \eqn{m^2 \ll (3H / 2)^2}), 
the solution to \Eq{psi-eom} is%
\begin{align}
\dot{\psi}(t) = -\fr{m^2}{3H} \psi(t) 
= \pm \mu \La \exp \biggl[ - \fr{m^2}{3H} t \biggr]  \,.
\end{align}

We can now check the assumption that the energy density of the $\tha$-$\psi$ 
system is dominated by $V_0$. The kinetic and potential energies of $\tha$ 
are well-approximated by $0$ and $V_0$, respectively, since $\tha$ and 
$\dot{\tha}$ approach zero exponentially at a rate of \eqn{\cO(H)}, as 
$\tha$ is not a slow-roll field (\eqn{\mu^2 \gg H^2)}. On the other hand, the 
kinetic energy of the slowly-rolling $\psi$ is subdominant to its potential energy, \eqn{m^2 \psi^2 / 2}. Thus, it suffices to demand 
\eqn{m^2 \psi^2 / 2 \ll V_0} at an earlier time when the length scale corresponding to the CMB 
epoch left the horizon, as \eqn{|\psi|} decreases with time.
Denoting by $N_\cmb$ the 
number of e-folds required to solve the horizon and flatness problems for that 
length scale, the requirement that \eqn{m^2 \psi^2 / 2 \ll V_0} at \eqn{t = - 
N_\cmb / H} yields the condition%
\begin{align}
1 \gg \fr{m^2 \psi^2}{2 V_0}
&= 
\fr{1}{V_0} \fr{9H^2 \mu^2 \La^2}{2m^2} 
\exp \biggl[ \fr{2m^2}{3H^2} N_\cmb \biggr]  \nn\\
&=
\fr{3 \mu^2 \La^2}{2m^2 \Mpl^2} 
\exp \biggl[ \fr{2m^2}{3H^2} N_\cmb \biggr]
\,.\eql{psi-potential-negligible}
\end{align}
Since \eqn{\mu^2 \gg m^2}, this implies%
\begin{align}
\La \ll \Mpl
\,.\eql{Lambda-small}
\end{align}

The seemingly innocuous relation~\eq{Lambda-small} is of fundamental 
importance. Nonperturbative quantum gravitational effects, such as virtual black 
hole formation, generically break all global symmetries~\cite{Kallosh:1995hi, 
Banks:2010zn}.  Fortunately, \eq{Lambda-small} ensures that gravity is weakly 
coupled at the cutoff, thereby exponentially suppressing such effects at the scale where 
the effective theory is defined, and protecting the global \eqn{\psi \to \psi + 
c} symmetry of our model.%
\footnote{Chaotic inflation~\cite{Linde:1983gd} with soft shift symmetry 
breaking by \eqn{m^2 \psi^2} can also be protected from nonperturbative 
gravitational effects if we lower the cutoff from $\Mpl$ to $\La$ by adding a 
shift-symmetric, nonrenormalizable interaction \eqn{(\del\psi)^4 / \La^4}. Even 
without lowering the cutoff, it is safe from perturbative gravitational effects 
for the same reason explained below.}

Perturbative graviton loops also violate the shift symmetry, but they can also 
be fully under control. Recall that we required, in the \eqn{\Mpl \to \infty} 
limit, that the \eqn{m^2 \psi^2} term be the only shift-symmetry breaking term 
in~\eq{lagrangian}. We also demanded that $\cL_\text{ct}$ 
contain only counterterms required by renormalization. Therefore, in perturbation theory,
all shift-symmetry violations ultimately stem from \eqn{\sqrt{-g}\, m^2 \psi^2}, so 
that every factor of $\psi^2$ without a derivative acting on it must be 
accompanied by a factor of \eqn{m^2 / \Mpl^2}. Thus, the condition for graviton loops not to spoil the softly broken shift symmetry is 
\begin{align}
1 \gg \fr{m^2 \psi^2}{\Mpl^2 \La^2} 
= \fr{m^2 \psi^2}{2V_0} \fr{\La^2}{\Mpl^2} \fr{2V_0}{\La^4} 
\,,\eql{graviton-loops-negligible}
\end{align}
when the CMB length scale left the horizon. 
This condition is already satisfied because of~\eq{psi-potential-negligible}, 
\eq{Lambda-small}, and the tree level relation%
\begin{align}
V_0 = \fr{\mu^4}{4\la} = \fr{\mu^2 \vev{\tha}^2}{4} \ll \La^4 
\,,\eql{V_0}
\end{align}
which follows from \eq{vev-less-than-cutoff} and the obvious requirement that 
\eqn{\mu^2 \ll \La^2}. 

The last theoretical consistency check is whether $\cL_\text{ct}$ is under 
control as an expansion in powers of \eqn{(\del \psi)^2 / \La^4}. This amounts to 
demanding \eqn{\dot{\psi}^2 / \La^4 \ll 1} when the CMB length scale left the 
horizon, which gives%
\begin{align}
\fr{\mu^2}{\La^2} \exp \biggl[ \fr{2m^2}{3H^2} N_\cmb \biggr] \ll 1
\,.\eql{psi-dot-smaller-than-cutoff}
\end{align}
Note that this is stronger than the condition~\eqn{\mu^2 \ll \La^2}.

%%%%%%%%%%%%%%%%%%%%%%%
\section{Phenomenology}
%%%%%%%%%%%%%%%%%%%%%%%

We now study the phenomenology of the simplest model~\eq{lagrangian}. 
Since primordial perturbations are only generated by degrees of freedom 
lighter than $H$, the dimensionless scalar power spectrum $\De_\cR^2$ does not 
depend on the heavy field $\tha$. Therefore, just as in single-field slow-roll 
inflation, we have%
\begin{align}
\De_{\cR}^2 \bigr|_\cmb
&= 
\biggl( \fr{H^2}{2\pi \dot{\psi}} \biggr)^{\!\! 2} \biggr|_\cmb  \nn\\
&=
\fr{H^4}{4\pi^2 \mu^2 \La^2} \exp \biggl[ -\fr{2m^2}{3H^2} N_\cmb \biggr]
\,,\eql{delta-rho-over-rho}
\end{align}
for which the 7-year WMAP data~\cite{Komatsu:2010fb} gives%
\begin{align}
\De_{\cR}^2 \bigr|_\cmb = (2.43 \pm 0.11) \times 10^{-9}
\,.
\end{align}
Similarly, the tilt of the scalar power spectrum, \eqn{n_\text{s} - 1}, can be 
computed as in single-field inflation by \eqn{n_\text{s} -1 = (2\eta - 6\vep) 
\bigr|_\cmb}, where the slow-roll parameters for our model are given by%
\begin{align}
\vep
= \fr12 \biggl( \fr{\Mpl \, m^2 \psi}{V_0} \biggr)^{\!\! 2}
\>,\quad
\eta
= \fr{\Mpl^2 m^2}{V_0}
\,.
\end{align}
Since $\eta$ is constant and \eqn{\vep / \eta = (m^2 \psi^2 / 2) / V_0 \ll 1} 
due to~\eq{psi-potential-negligible}, 
we obtain%
\begin{align}
n_\text{s} - 1
\simeq 
2\eta = \fr{2m^2}{3H^2}
\,.\eql{n-1}
\end{align}
This should be compared to the 
current constraint~\cite{Komatsu:2010fb}, \eqn{n_\text{s} - 1 = -0.032 \pm 
0.012} (68\% CL\@). To be within $3\sigma$ of the central value, let us impose 
\eqn{\eta < 2 \times 10^{-3}}. This then ensures that a sufficient number of 
e-folds can easily be accommodated, since the exponential factor 
in~\eq{psi-potential-negligible} and~\eq{psi-dot-smaller-than-cutoff} is 
\eqn{\simeq 1} for $N_\cmb$ of \eqn{\cO(10^2)}.   

To see how all the theoretical consistency conditions are satisfied, 
we express them in terms 
of $\mu^2$, $\La^2$, $\eta$, $\De_\cR^2$, and $r$. To eliminate the dependence 
of our expressions on $N_\cmb$, $m^2$, and $H^2$, we use the 
relations~\eq{delta-rho-over-rho}, \eq{n-1}, and% 
\begin{align}
r = 16 \vep |_\cmb = \fr{2}{\pi^2 \De_\cR^2} \fr{H^2}{\Mpl^2} 
\,,\eql{r}
\end{align}
respectively. The condition~\eq{psi-potential-negligible} implies an upper 
bound on the tensor-to-scalar ratio $r$:%
\begin{align}
r \ll 16 \eta < 3.2 \times 10^{-2}
\,.\eql{r-bound}
\end{align}
The current experimental bound is \eqn{r < 0.20} at 95\% 
CL~\cite{Komatsu:2010fb}. 
On the other hand, the condition~\eq{vev-less-than-cutoff} becomes%
\begin{align}
1 
&\gg 6\pi^2 \De_\cR^2 r \fr{\Mpl^4}{\mu^2 \La^2}  \nn\\
&\simeq \fr{\De^2_\cR}{2.43 \times 10^{-9}} \fr{r}{3.2 \times 10^{-2}}
\biggl( \fr{0.03}{\mu / \La} \biggr)^{\!\! 2}
\biggl( \fr{0.05}{\La / \Mpl} \biggr)^{\!\! 4}
\,.
\end{align}
Combined with the bound on $r$ from~\eq{r-bound}, this implies that the 
condition~\eq{vev-less-than-cutoff} can be satisfied with large separations 
among $\mu$, $\La$, and $\Mpl$, thus keeping the effective theory under control 
as well as suppressing nonperturbative gravitational effects. 

The rest of the consistency conditions are easy to satisfy. 
The condition~\eq{psi-dot-smaller-than-cutoff} becomes% 
\begin{align}
1 
&\gg \fr{\pi^2 \De_\cR^2 \, r^2}{16} \fr{\Mpl^4}{\La^4}   \\
&=2 \times 10^{-7} 
\fr{\De_\cR^2}{2.43 \times 10^{-9}}
\biggl( \fr{r}{3.2 \times 10^{-2}} \biggr)^{\!\! 2}
\biggl( \fr{0.05}{\La / \Mpl} \biggr)^{\!\! 4}
\,,\nn
\end{align}
ensuring that the derivative expansion in \eqn{(\del \psi)^2} is well under 
control. The slow-roll condition \eqn{m^2 \ll (3H / 2)^2} is automatically 
satisfied, as \eqn{m^2 / (3H / 2)^2 = 4\eta / 3 \ll 1}. Finally, 
the condition \eqn{\mu^2 \gg (3H / 2)^2} becomes%
\begin{align}
1 
&\gg \fr{9\pi^2 \De_\cR^2 r}{8} \fr{\La^2}{\mu^2} \fr{\Mpl^2}{\La^2}  \\
&= 4 \times 10^{-4} \fr{\De_\cR^2}{2.43 \times 10^{-9}}
\fr{r}{3.2 \times 10^{-2}} 
\biggl( \fr{0.03}{\mu / \La} \biggr)^{\!\! 2}
\biggl( \fr{0.05}{\La / \Mpl} \biggr)^{\!\! 2}
\,,\nn
\end{align}
which is robustly satisfied.

%%%%%%%%%%%%%%%%%%%%%%%%%%%%%%%%%%%%%%%%%%
\section{Discussion and Outlook}
%%%%%%%%%%%%%%%%%%%%%%%%%%%%%%%%%%%%%%%%%%

We have proposed a new framework, gyroscopic inflation, in which the stability 
of the field driving the expansion is controlled by a slow-roll field 
with a shift symmetry that is only softly broken. 
We have formulated the simplest realization and
examined the conditions for it to be a consistent effective field theory. 
We have seen that those consistency conditions, combined with observational 
constraints, lead to the prediction that the tensor-to-scalar ratio for 
primordial perturbations should be $r \ll 3 \times 10^{-2}$. Detection of 
primordial gravitational waves may therefore be within reach of the next 
generation of CMB experiments, \eqn{r \gsim \cO(10^{-2})}. 

On the other hand, the simplest consistent effective field theory of inflation 
with a softly broken shift symmetry is \eqn{m^2 \psi^2} chaotic 
inflation~\cite{Linde:1983gd}, 
which predicts \eqn{r = 4 (1 - n_\text{s}) = 0.13} for the current central 
value of \eqn{1 - n_\text{s} = 0.032}. It is therefore possible that this model 
could be ruled out by future CMB experiments. Then, divorcing $r$ from 
\eqn{1 - n_\text{s}} would require introducing a second field 
that dominates the potential energy.  Preserving the shift symmetry would then
lead to gyroscopic inflation.
 
Like hybrid inflation \cite{Linde:1993cn}, the minimal gyroscopic model has a tension with the 
measurement of 
\eqn{n_\text{s} - 1}, although not so severe as to exclude it. Improving this 
would require an additional degree of freedom with mass \eqn{\ll H}. The 
model-building challenge would be to incorporate such an extension as an 
integral part of gyroscopic inflation, e.g., by embedding $\psi$ and the new 
field into a single multiplet.

In both the minimal and improved models, non-Gaussianity in primordial 
perturbations deserves further investigation. In gyroscopic inflation, in 
contrast to conventional slow-roll inflation, a nearly constant potential does 
not necessarily imply that the slow-roll field $\psi$ is almost free, as it has 
derivative interactions. Therefore, non-Gaussianity should be present at some 
level even in the simplest model, and could possibly be more pronounced in the 
improved models.  Furthermore, in an improved model, the existence of additional light degrees of 
freedom would generically lead to non-adiabaticity in the primordial 
perturbations.
 
Another reason to modify the simplest model relates to 
the issue of ultraviolet (UV) completion of the effective 
theory~\eq{lagrangian}. One 
might think that the softly broken shift symmetry for $\psi$ should indicate 
that $\psi$ is the pseudo-Nambu-Goldstone boson (pNGB) of a broken 
symmetry. The problem is that the value of $\psi$ during inflation is larger 
than $\La$, and hence larger than its decay constant. For example, at the end of 
inflation we have \eqn{|\psi| / \La = 3H |\dot{\psi}| / (m^2 \La) = 3H \mu / m^2 
\gg 1}. As in $m^2 \psi^2$ chaotic inflation, this does not invalidate the 
\emph{effective} theory, since the shift symmetry is only softly broken and the 
shift-symmetry breaking graviton loops are negligible, as we checked in 
\eq{graviton-loops-negligible}. However, a pNGB field value much larger than its 
decay constant is possible only if the broken symmetry group is non-compact 
(e.g., $\R$ rather than U(1)). This is believed to be incompatible 
with quantum gravity~\cite{Banks:2010zn}. One possible resolution may be that a 
large field value effectively arises as a collective effect, as in 
$N$-flation~\cite{Dimopoulos:2005ac}. Another possibility is that the 
noncompactness of $\psi$ could be related to flat directions in a supersymmetric 
theory. 

To summarize, both the measurement of \eqn{n_\text{s} - 1} and the consistency 
of the UV completion motivate introducing additional light degrees of freedom. 
Non-Gaussianity and non-adiabaticity may then probe how the 
minimal model of gyroscopic inflation is extended. 

This work was supported by the DOE grant DE-FG02-97ER41022.

%%%%%%%%%%%%%%%%%%%%%%%%%%%%%%%%%%%%%%%%%%%%%%%%%%%%%%%%%%%%%%%%%%%%%%%%%%%%%%%%
%%%%%%%%%%%%%%%%%%%%%%%%%%%%%%%%%%%%%%%%%%%%%%%%%%%%%%%%%%%%%%%%%%%%%%%%%%%%%%%%

\end{document}